# Young people's perceptions and recommendations for conversational generative artificial intelligence in youth mental health

Adam Poulsen[1, *, https://orcid.org/0000-0002-0001-3894]
Ian B Hickie[1, https://orcid.org/0000-0001-8832-9895]
Carla Gorban[1, https://orcid.org/0009-0003-9568-1792]
Zsofi de Haan[1, https://orcid.org/0009-0003-3341-5982]
William Capon[1, https://orcid.org/0000-0001-6500-9629]
Ebenezer Eyeson-Annan[2, https://orcid.org/0009-0009-3526-0201]
Jalal Radwan[2, https://orcid.org/0009-0005-2929-7252]
Elizabeth M Scott[1, https://orcid.org/0000-0003-3907-0324]
Frank Iorfino[1, †, https://orcid.org/0000-0003-1109-0972]
Haley M LaMonica[1, †, https://orcid.org/0000-0002-6563-5467]

[1]Brain and Mind Centre, The University of Sydney. Sydney, Australia.
[2]Uncapt. Sydney, Australia.
[†]Joint senior authors

**Corresponding author:**

[*]Adam Poulsen. 94 Mallett St, Camperdown 2050, Australia. adam.poulsen@sydney.edu.au

## Abstract

Conversational generative artificial intelligence agents (or genAI chatbots) could benefit youth mental health, yet young people's perspectives remain underexplored. We examined the Mental health Intelligence Agent (Mia), a genAI chatbot originally designed for professionals in Australian youth services. Following co-design, 32 young people participated in online workshops exploring their perceptions of genAI chatbots in youth mental health and to develop recommendations for reconceptualising Mia for consumers and integrating it into services. Four themes were developed: (1) Humanising AI without dehumanising care, (2) I need to know what's under the hood, (3) Right tool, right place, right time?, and (4) Making it mine on safe ground. This study offers insights into young people's attitudes, needs, and requirements regarding genAI chatbots in youth mental health, with key implications for service integration. Additionally, by co-designing system requirements, this work informs the ethics, design, development, implementation, and governance of genAI chatbots in youth mental health contexts.

**Keywords:** Mental health, Youth, Artificial intelligence, Chatbot, Co-design workshop, Prototyping, User experience

## 1. Introduction

Youth mental health is in crisis[1]. Worldwide, young people aged 12-25 are experiencing high and, in most cases, rising rates of anxiety, depressive symptoms, psychological distress, self-harm, suicide, and other mood disorders[2,3]. Contributing factors span psychological, social, and economic domains, creating a multifaceted challenge[4]. The wide spectrum of mental illnesses in presenting populations adds to the challenge of addressing all clients' needs, risking the exclusion of some from care access[5]. Socioeconomic, cultural, and environmental factors compound the issue, adding stressors such as social stigma, employment and housing insecurity, intergenerational inequality, especially in low- and middle-income countries[1,2]. Worsening the problem, services are overburdened[1] and suffer substantial gaps in workforce, service data, regulation, service specificity, coverage, and coordination[2,6]. Prevention and early, effective, and high-quality intervention of emerging mental illnesses at their peak onset of 15 years is essential to positively change the daily lives and future life-course of young people[1]. Yet, the state of youth mental health often leads to delayed, reactive, and inappropriate treatment decisions [5,7].

The digitisation of youth mental health shows promise in supporting services and stakeholders and alleviating the present issues[8]. This includes the integration of various digital mental health technologies (DMHTs), including smartphone apps, videoconferencing platforms, digital phenotyping, virtual and augmented reality, and artificial intelligence (AI) systems[9]. Principally, DMHTs may help to optimise services with an improved, flexible, technology-driven user experience, generate actionable and useful data for mental health service providers and professionals to deliver highly personalised and measurement-based care, and provide new avenues for prevention, delivery, and treatment support through DMHT interventions[9]. Research indicates that the DMHTs support shared and informed decision-making by enhancing the breadth of information considered to best treat the myriad mental illnesses in presenting populations in youth mental health services[10,11].

Advances in large language models (LLMs) and machine learning techniques present an opportunity for introducing conversational generative AI agents (or genAI chatbots) in youth mental health to engage users, classify large volumes of text, answer questions, translate languages, and generate new content based on user input beyond standardised, untailored responses of conversational AI alone[12], benefiting mental health professionals and clients[13,14]. For professionals, research highlights that conversational genAI in mental health is positioned to increase efficiency to clinician workflows by supporting assessment[15], assisting in developing empathic clinician documentation and client-facing interactions, and reducing time spent on administrative clinician tasks, thus recovering critical time with clients[14]. For clients, research indicates that conversational genAI in this context improves mental health outcomes[16,17], reduces symptoms[18], benefits psychoeducation and self-adherence[19], and user satisfaction[19], and supports health information comprehension and informed decision-making for non-experts[20]. Going forward, conversational genAI is expected to play a significant role in supporting accessible, scaleable, and personalised mental healthcare, owing to its considerable context-driven adaptability and data collection and processing capabilities[21-23]. This aligns with the existing trend of young people turning to publicly accessible, general purpose systems, such as OpenAI's ChatGPT, for mental health support[24]. Yet, these tools raise key questions around trust, usability, effectiveness, engagement, risk of harm, and inclusion of vulnerable groups[16,25-29]. This calls for meaningful co-design with users and other stakeholders to develop a rich understanding of their needs and preferences, along with the associated ethical, design, development, implementation, and governance priorities.

As is typical in digital health research, much of the literature focuses on evaluating technologies for functionality and effectiveness, risking obfuscation of qualitative insights concerning user perceptions, user experience, and human-computer interaction and human-centred design considerations[30]. Furthermore, potential end users are often only involved in the final stages of technology development, excluding an account of their perceptions, needs, and characteristics in the early stages[31]. A scoping review by Balan, et al. [32] on conversational AI in youth mental health noted this trend, finding that input from potential end users during the design and development stages is often overlooked. There is limited research focusing on examining end user perceptions and recommendations relating to the features, scope, content, personality, ethics, and subsequent design considerations to inform the development of conversational AI (and genAI[33]) in this space[32,34].

To fill this gap, this paper reports the co-design findings from developing a genAI chatbot, the Mental health Intelligence Agent (Mia)[35], for enhancing triage, clinical assessment, formulation, treatment planning, and self-management in youth mental health services and examines young people's relevant perceptions.

## 2. Results

### 2.1. Participant characteristics

A total of 32 young people participated in the workshops, among which 19 completed a follow-up survey reporting demographics (see Figure 1). They ranged in age from 18 to 30 years (mean age 25 years) and first sought mental health support at ages 9 to 23 years (mean first support-seeking age 18 years). The gender distribution was women (n=11), men (n=5), non-binary (n=1), agender (n=1), and not disclosed (n=1). They identified as women (n=11), men (n=5), non-binary (n=1), and agender (n = 1), with one opting not to disclose. Intersectional identities reported included lesbian, gay, bisexual, transgender, gender diverse, intersex and queer (n=8), culturally and linguistically diverse background (n=8), belonging to a religious and/or spiritual group (n=4), carer responsibility (n=1), neurodivergent (n=1), chronic illness (n=1), and none (n=4), with identities reported non-exclusively. The majority reported lived experience of mental health that had impacted their everyday lives (n=14), with others reporting no lived experience (n=4) or choosing not to respond (n=1). Just over half reported currently receiving treatment (n=10), while others reported not currently receiving treatment (n=6) or opted not to disclose their treatment status (n=2), with one choosing not to respond.

**Figure 1. Participant demographics**

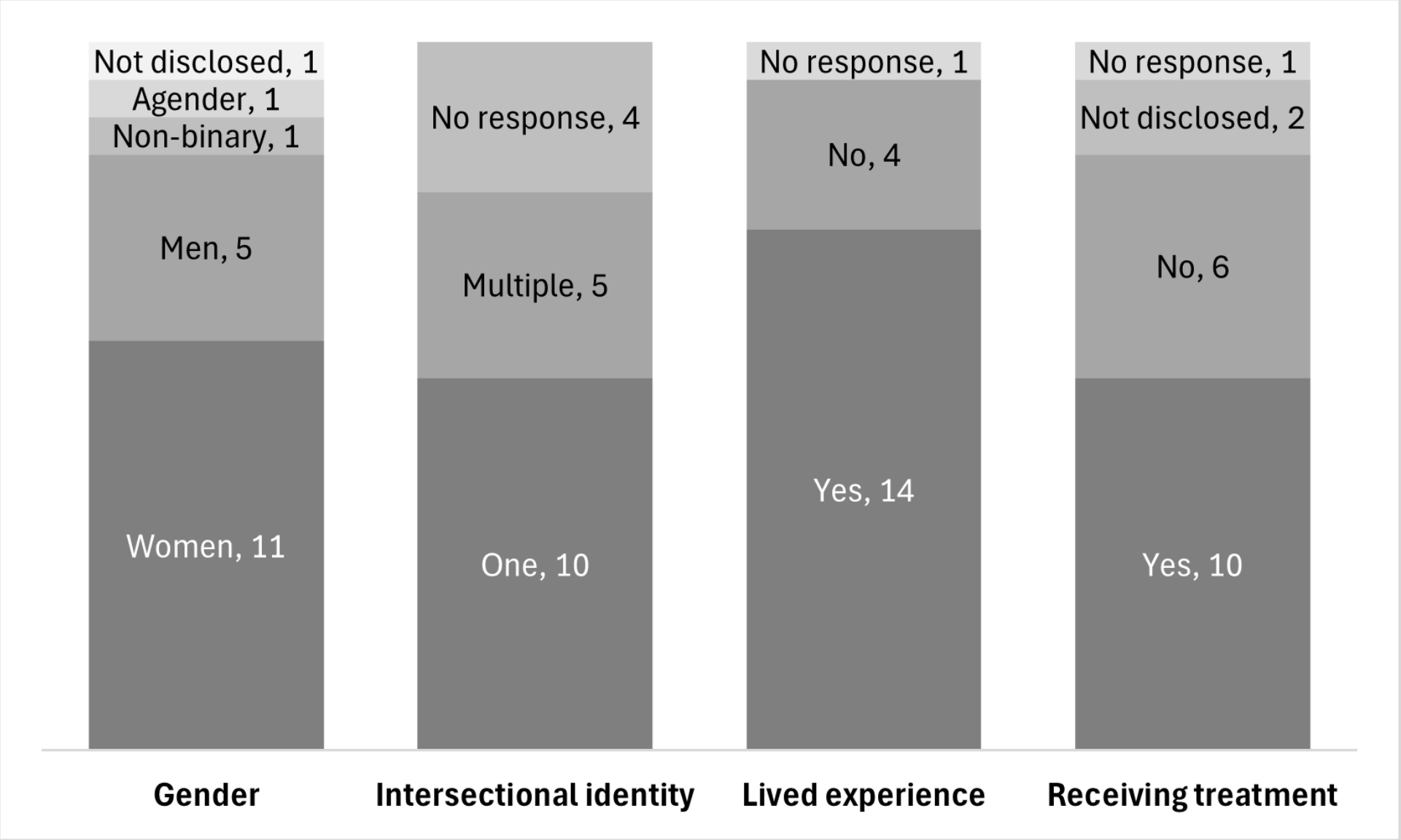

## 2.2. Reflexive thematic analysis

Following reflexive thematic analysis[36], four themes were developed from the data: (1) Humanising AI without dehumanising care, (2) I need to know what's under the hood, (3) Right tool, right place, right time?, and (4) Making it mine on safe ground. Figure 2 reports the thematic map[36], indicating the relationships between the themes.

**Figure 2. Thematic map, indicating the four themes identified in this study. Relationships between themes are indicated by directional lines and labels.**

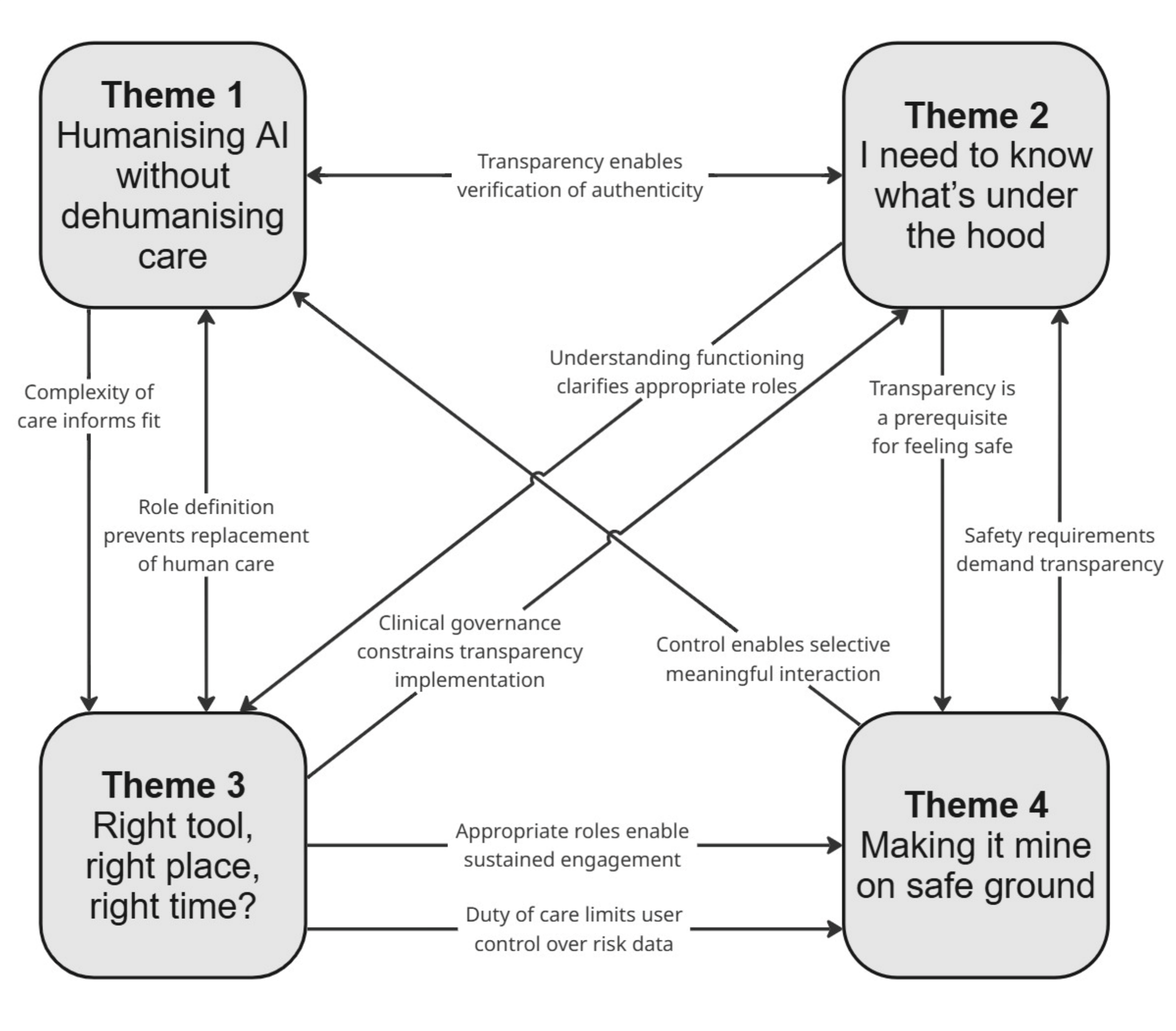

### 2.2.1. Theme 1: Humanising AI without dehumanising care

This theme speaks to young people's perceptions of introducing genAI chatbots into the fundamentally *human* mental healthcare environment, and the necessary harmonisation between using advanced systems while preserving the centrality of humans in the delivery of care. Young people discussed the need to develop genAI chatbots equipped with human-like qualities to echo empathy, while ensuring that these tools support rather than replace clinicians.

> *'It can't outright replace actual psychologists and psychiatrists, because the human connection is still important at the end of the day, it's not something that can be replicated by AI or chatbots' (YP15)*

Young people's central concern about preserving human connection was driven by doubts about whether genAI chatbots could truly comprehend the complexity of mental health experiences. The profoundly personal, contextual, and nuanced nature of mental health was felt to exceed what genAI chatbots could understand, raising apprehension about them replacing *real* care and creating a sense of alienation and loss of human touch.

> *'I feel a lot of the problems that I experience as a young person with lived experience are a lot more complex than maybe the advice that a system like this might be able to offer' (TP7)*

These concerns were compounded by more fundamental fears about genAI chatbot capabilities and their potential to dehumanise care through unintended consequences. These included genAI hallucinations and poor sustainability of genAI, as well as existing problems in health care that are at risk of being exacerbated with the introduction of genAI, such as poor information recall, risk of bias in decision-making, fostering unhealthy dependence or inappropriate attachment, and perpetuation of existing social stigma, leading to feelings of judgement. In this way, young people conveyed that the perceived cost of humanising genAI chatbots must not come at the expense of eroding the human foundation of mental healthcare, or the human experience generally. Additionally, they emphasised the need for ongoing AI ethics engagement and for considering a diversity of stakeholder perspectives beyond lived experience of mental health in development processes to ensure these tools serve all young people appropriately and best advance the tailoring of user experiences and delivery of personalised outputs rather than a generic approach, necessitating a broad and diverse training data set and application of ethical frameworks.

During user testing, which utilised Mia (see Figure 3), some young people tested these concerns by challenging the genAI chatbot with difficult queries, complex or vague mental health experiences and concerns, and non-English language inputs in an effort to *break* it. This scepticism, as well as comparisons made to other tools like ChatGPT, reaffirmed young people's doubts about whether current genAI chatbots (either general or purpose-built) are meaningful and sophisticated enough for mental healthcare.

**Figure 3. Original Mental health Intelligence Agent (Mia) user interface intended for health professionals, utilised for early stages of user testing. [SEE END OF DOCUMENT FOR FIGURE 3]**

Yet, young people also recognised the importance of designing genAI chatbots to reflect human-like empathy and acknowledged the potential value in thoughtfully developed tools. This was especially important for building rapport, trust, and credibility, evidencing their value. Primarily, this was expected to be achieved through genAI chatbots being equipped with key language capabilities. Those included, having an emotionally resonant, supportive, and personable tone, using accessible language that communicates mental health information without sounding clinical, nuanced understanding of youth and mental health related language (including multilingual capabilities), ability to sensitively address confronting topics like suicide and death, adaptive communication tailored to individual users, and clear prompts to elicit key mental health information from users without pressuring them for responses. Beyond language, young people also emphasised that human-like empathy in genAI chatbots ought to be realised through high standards for information recall, case formulation accuracy, and intersectionality sensitivity.

> *'Obviously you want the chatbot to be professional, but you also don't want them to be lacking of empathy' (YP19)*

#### 2.2.2. Theme 2: I need to know what's under the hood

This theme captures young people's expectations of system transparency and accuracy in their interactions with genAI chatbots in youth mental health. Young people expected transparency across multiple dimensions, including being able to inspect a genAI chatbot's evidence base, examine its decision-making processes, and have clarity on how their data would be used and stored. Equally important were expectations about accuracy. Young people wanted genAI chatbots to interpret user inputs aptly, draw appropriate inferences from them, and provide reliable outputs grounded in evidence.

Based on user testing feedback, they valued functionality that provided insight into the inner workings of genAI chatbots and enabled traceability. Specifically, they appreciated that the genAI chatbot could explain its interpretations of user inputs plainly, identify its rationale for extracting and organising specific components of the user inputs, and detail the reasoning process from initial inferences about the user through to treatment, self-management, and service navigation recommendations, with each step linked to evidence-based knowledge from the research literature. With this information made visible, they felt positioned to make their own independent determinations about the accuracy of system interpretations, decisions, and outputs, rather than accepting the genAI chatbot's conclusions at face value.

> *'A lot of people say, "Oh, you can't trust AI 100%". And I'm sure you can't really trust anything 100% but just to show that it bases its answers on reasonable things, I think it's important to show its thought process' (YP14)*

Most young people noted that transparency and traceability are empowering, enabling greater agency over their mental health information, which in turn advances trust and credibility in genAI chatbots. However, some also identified a potential tension about being too transparent. That is, presenting AI-generated insights that suggest negative trajectories about one's mental health might undermine empowerment by triggering negative feelings and disengagement from the tool and mental health services more broadly, regardless of accuracy. This was especially a concern in a hypothetical discussion about genAI chatbots detecting risk of immediate risk factors, such as imminent physical violence or high rates of suicidal thoughts and behaviours. Despite these concerns, the prevailing view was that withholding information (either positive or negative) from users by design and stifling system transparency is intrinsically disempowering and unethical.

> *'I feel like information that's available to clinicians...like the point of a clinician is to be explaining stuff to you, right? So, it doesn't really make sense for the info [sic] to be gate-kept like that. I feel I would be a bit annoyed if I found out that there was a clinician's view on this and I didn't have access to all of it' (YP28)*

Young people offered key insights into how transparency and accuracy expectations could be better established within system design. To advance transparency, they recommended addressing these requirements at the onboarding stage through an explanatory introduction detailing the genAI chatbot's functions, data practices, reasoning processes, and underlying architecture. Clear communication about system operations from initial contact was viewed as essential for setting appropriate expectations and establishing a foundation for trust (see Figure 4 for revisions made to Mia in response to young people's feedback). To enhance accuracy, they recommended scaling the knowledge base of genAI chatbots to incorporate global research insights and broader and more diverse data sources and variables.

**Figure 4. Revised Mental health Intelligence Agent (Mia) user interface updated for young people based on feedback, used for later stages of user testing. [SEE END OF DOCUMENT FOR FIGURE 4]**

Young people also assessed genAI chatbots against their existing experiences with DMHTs (e.g., mental health apps) and other technologies used for mental health support, including social media platforms and general-purpose AI tools (e.g., ChatGPT). These experiences shaped their expectations. They shared that digital tools could offer accessibility and anonymity while also posing risks, including misinformation, echo chambers, and superficial engagement. In contrast, they expected genAI chatbots in youth mental health to meet higher standards of both transparency and accuracy than unregulated information and tools operating outside of mental health service environments, demonstrating clear advantages in terms of accuracy, evidence-grounding, and appropriate safeguards communicated directly to users.

> *'In terms of accuracy, it's pretty important to me that it is good. Otherwise, it's pretty worthless [...] Just the retention of information from an LLM is just not at the same level, or even anywhere close to, obviously, the clinician's ability to interpret and understand and remember back to previous conversations and have context' (YP12)*

### 2.2.3. Theme 3: Right tool, right place, right time?

This theme addresses young people's fundamental questions and reactions about the *fit* of genAI chatbots in youth mental health. It examines whether genAI chatbots represent the right tool for the job, positioned at the right touchpoints within the service ecosystem, and available at the right times in young people's care journeys. Young people discussed how genAI chatbots might fit within the existing multi-stakeholder sociotechnical system of mental healthcare, what roles they should appropriately fill, where young people would encounter them, and what value those interactions would have. They also explored how these chatbots could be integrated into established services, practices, and relationships without being disruptive. Important framing here is that the discussions focused on service integration, scoping the conversation to explore genAI chatbots that are made accessible only to young people who were seeking care or actively engaged in care at a service, ranging from initial contact with the mental health system to being a regular client.

Much of the discussion centred on identifying the potential roles that genAI chatbots could play in youth mental health. First, serving as *navigators* to help young people identify, understand, and access the right mental health services. Here, providing access means providing service-approved contact information or website links to young people for services applicable to their needs. Some young people noted that these recommendations should be categorised by location and cost, enabling users to make decisions aligned with their preferences and needs, but without requiring additional potentially invasive personal data collection, such as location sharing. Second, as assessment tools (or *assessors*) to iteratively interpret user input concerning a young person's mental health information (e.g., previous and current symptoms, case formulation, support, and service engagement) and other associated contextual factors, elicit information relating to critical signs to assess for in young people, develop summaries detailing a young person's needs, and generate highly personalised and context-driven treatment, self-management, and service navigation recommendations ongoingly. Third, genAI chatbots as mental health *educators* to detail and compare different treatments, outline the steps involved in engaging with a service or treatment, inform individuals about what to expect at different service engagement stages, and communicate the research evidence about particular mental health conditions, treatments, and self-management strategies, ultimately benefiting mental health literacy and supporting informed decision-making. Other minor roles briefly mentioned included supporting the creation of personalised support plans and safety plans, serving as informal spaces where young people can vent or seek advice, and facilitating some administrative processes such as booking appointments.

Young people considered the touchpoints where users would encounter genAI chatbots within their care journeys. Key touchpoints included initial contact with a service (or pre-intake), intake processes, preparing for clinical sessions, situations requiring instant support, and ongoing service engagement (any point after intake). From these two dimensions (i.e., roles and touchpoints), specific use cases emerged that illustrated how navigation, assessment, and education could be enacted at different moments in young people's engagement with genAI chatbots in youth mental health (see Table 1). While certain use cases are particularly linked to specific touchpoints, such as self-screening at pre-intake, completing an initial assessment during intake, or accessing emergency contacts when in crisis, some use cases are ongoing and not bound to particular touchpoints. Young people expressed that it was important for users to be able to access psychoeducational content, ask questions about treatment, self-management, and service navigation recommendations and ideate alternatives, receive updated recommendations based on current needs, review personal information, and use mental health tracking functions at any point after intake. Table 1 details the use cases organised by the touchpoints where they are most salient, while recognising that the chatbot's utility extends beyond these discrete moments (see the ongoing service engagement touchpoint).

**Table 1: Use cases for genAI chatbots in youth mental health emerging from the workshop discussions, organised by touchpoints in the care journey.**

| Touchpoint | Use case |
|---|---|
| Pre-intake | • Complete self-screening to understand if help is needed (assessor)<br>• Explore service options based on needs without commitment (navigator and assessor)<br>• Explore care options based on needs without commitment (assessor and educator) |
| Intake | • Get orientation to the service based on needs (navigator and educator)<br>• Book initial appointment (navigator)<br>• Complete intake assessments and forms (e.g., service registration forms or consent for treatment) (assessor and navigator)<br>• Access content about the intake process and what information is needed (educator)<br>• Create initial recommended treatment plan based on needs (assessor)<br>• Create initial safety plan (assessor and navigator) |
| Preparing for clinical sessions | • Receive reminders about appointments (navigator)<br>• Receive reminders about therapy homework (navigator)<br>• Access pre-session preparation materials (educator and navigator) |
| Moments of crisis | • Get orientation to the appropriate level of care based on risk assessment (assessor and navigator)<br>• Receive assessment-based, distress management recommendations (e.g., self-care and coping strategies) (assessor and educator)<br>• Access emergency contacts or crisis service information (navigator)<br>• Document crisis episodes for later review with the clinician (assessor) |
| Ongoing service engagement | • Access psychoeducational content about mental health concerns (educator)<br>• Query about treatment, self-management, and service navigation recommendations (educator and navigator)<br>• Receive updated personalised recommendations based on current needs (assessor and educator)<br>• Review personal information (e.g., self-care recommendations, safety plans, timeline, tracked data) (assessor and educator)<br>• Track symptoms, mood, or other metrics (assessor) |

Young people reflected on broader issues within the mental health system, including access, efficiency, and quality of care, and raised a key consideration about whether genAI chatbots could address these issues. They noted that genAI chatbots could potentially improve access to support and care by supporting intake processes, managing demand on overstretched services, reducing wait times, and enabling a soft entry into care for hesitant young people who may find traditional service engagement difficult for a variety of reasons, such as stigma, privacy, and power imbalance concerns. The option to interact with a non-judgmental system before encountering mental health professionals was viewed as particularly valuable for reducing barriers to help-seeking. Additionally, genAI chatbots could enhance care efficiency by automating routine tasks and improving user data retention and sharing, thereby reducing the burden of repeatedly retelling one's historical and current mental health experiences and concerns across multiple service touchpoints, which is a common source of frustration and re-traumatisation for young people navigating fragmented care systems.

When discussing what would make interactions with genAI chatbots genuinely useful at various touchpoints, young people emphasised the importance of personalised and actionable treatment, self-management, and service navigation recommendations. They wanted genAI chatbots to provide recommendations tailored to individual circumstances. Young people wanted to see treatment and self-management options that would require implementation with a clinician (e.g., cognitive behavioural therapy) and those centred around self-care that young people could implement independently between formal care contacts (e.g., mindfulness). Moreover, they expected that both types of care options would be tailored to them based on their previous experiences with mental healthcare, offering alternatives to previously unsuccessful treatments. This requirement positioned genAI chatbots as tools that could extend the reach of professional care and empower young people to take active roles in managing their mental health. However, young people also identified the potential risk of conflicting recommendations between genAI chatbots and clinicians that could undermine trust in both the technology and professional care, highlighting the need for careful integration and alignment with clinical practice.

#### 2.2.4. Theme 4: Making it mine on safe ground

This theme captures what enables young people to sustain engagement with genAI chatbots in youth mental health, with a particular emphasis on the dual requirements of choice and safety. Young people highlighted that meaningful engagement depends on the capacity to customise interactions to suit their individual preferences while feeling confident that they are protected, both in terms of data privacy and security and appropriate system safeguards to ensure safe interactions. Without choice, the genAI chatbots are at risk of feeling non-personalised, restrictive, overbearing, and inflexible. Without safety, they are susceptible to being perceived as risky and untrustworthy. Both elements were expressed as prerequisites for young people to invest in using these tools over time.

Young people emphasised the importance of maximising choices about how they interact with genAI chatbots wherever possible and appropriate. This included preferences about interaction modalities (e.g., text versus voice input), access points (e.g., mobile app versus web app), the structure of conversations (e.g., guided by prompts or open-ended), and the pacing of engagement (e.g., quick check-ins versus extended interactions). This trend resonated throughout all the workshops, with young people making it clear that one size does not fit all. Individuals have varying and evolving communication preferences, access modalities, comfort levels, and needs at different touchpoints and with genAI chatbots serving different roles, so providing options to allow users to tailor the experience was suggested to increase the likelihood of sustained use.

Related to choice was the question of how much information genAI chatbots in youth mental health should provide in responses. Young people had diverse preferences about the level of detail they wanted. Again, echoing the importance of choice, some appreciated comprehensive explanations while others found that too much information is overwhelming. The majority expressed a preference for an opt-in approach where concise responses are provided by default, and users can choose to request more detailed information. This approach respects that information needs vary by individual, context, and topic, and empowers users to control the depth of engagement rather than having genAI chatbots (or designers) dictate it.

Young people also discussed practical questions about how genAI chatbots should be accessed technically. There was an emerging preference for access via mobile apps, recognising that young people primarily use smartphones for digital interactions and that apps provide a more seamless, familiar, and on-demand experience compared to web browser-based platforms. Additional questions arose regarding whether users need to register personal accounts with identifying information, whether genAI chatbots require internet connectivity, and how accessibility features should be incorporated for users with diverse needs. These technical access considerations impact whether young people can meaningfully engage with genAI chatbots in the way they choose.

Safety was considered non-negotiable for engagement, with young people expressing concerns across two distinct but interconnected dimensions: data safety and interaction safety. On data safety, young people wanted clarity on what data would be collected, how it would be stored to maintain privacy and security, who would have access to it, how long it would be retained, and whether it would be used to retrain genAI chatbot models. Beyond organisational data security measures and privacy policies to ensure data safety, young people wanted user-side control over their own data and privacy. A critical aspect of this control concerned protecting the privacy of their raw data (e.g., user inputs). Young people emphasised that for the sake of promoting unfiltered interactions and protecting privacy and autonomy, their raw data should remain private and under their control. They wanted to decide whether to share the raw data and with whom.

On interaction safety, young people emphasised that genAI chatbots should recognise and respond appropriately when a user is at risk. Young people expected these tools to detect when users discuss unsafe situations (e.g., sharing imminent plans of self-harm or violence) or when they identify patterns suggesting deteriorating mental health or elevated risk. When such risks are detected, young people suggested users should be directed to mental health professionals (e.g., online chat crisis support services or local emergency crisis support services).

Data safety and interaction safety intersect in complex ways. Young people's desire for data privacy (i.e., keeping raw data private) exists in tension with the need for interaction safety (i.e., ensuring clinicians can intervene when risk is detected). Young people resolved this tension by distinguishing between raw data and AI-generated insights. While raw conversation data should remain private, AI-generated insights developed from those interactions (i.e., summaries detailing a user's needs, personalised treatment, self-management, and service navigation recommendations, and deterioration or risk detection) should be accessible to clinicians and services. This distinction allows genAI chatbots to fulfill protective functions (monitoring for risk, flagging concerning patterns, enabling timely intervention, supporting coordinated care, and upholding the duty of care) without compromising the privacy of users' raw data. In this model, young people maintain a private space for authentic expression while clinicians receive the clinically relevant information needed to provide appropriate support and ensure safety.

> *'So in the clinician's view [...] if they can also see the recommendations that the clients going to get, that would be helpful for not only safety, but also for the clinician to know what their clients kind of do as well' (YP12)*

## 3. Discussion

This study engaged 32 young people in workshops to explore their perceptions of genAI chatbots in youth mental health and to develop recommendations for transforming Mia (originally a tool for health professionals) into a solution for consumers and integrating it into youth mental health services. Four themes were developed that, together, describe how these technologies might meaningfully support mental health, realise potential benefits, navigate possible risks, and fit within the broader socio-technical context. The themes address critical knowledge gaps regarding genAI chatbots for consumers in youth mental health. Namely, how to design human-like AI without devaluing human-delivered care (Theme 1), what transparency mechanisms are positioned to build trust in high-stakes mental health contexts (Theme 2), where and when genAI chatbots may add value across care journeys (Theme 3), and how to sustain engagement while balancing user choice and system safeguards through nuanced data governance (Theme 4).

Findings indicate that young people expected genAI chatbots in youth mental health to communicate empathetically while fearing that human-like genAI might inappropriately reduce access to human clinicians or be used to justify reduced access. This tension reflects longstanding ethical concerns about technologies in care settings mimicking human qualities to substitute rather than supplement human relationships[37]. Furthermore, it echoes perspectives emphasising that human empathy in caregiving cannot be fully replicated by advanced systems, as they lack the understanding of lived human experience necessary for authentic care[38,39]. Scepticism about whether genAI chatbots could comprehend the profoundly personal, contextual nature of mental health experiences is grounded in phenomenological understandings that lived experience resists algorithmic reduction[40]. Still, the enduring expectation for empathetic genAI chatbots presents a conflict that requires careful balancing. To address it, high standards for youth-centric language adaptation, information recall, case formulation accuracy, and intersectionality sensitivity were proposed as acceptable advances toward human-like empathy in genAI chatbots.

Subsequently, another critical tension emerged around performance standards. Young people in this study held genAI chatbots to higher standards than human clinicians, expecting exceptional system comprehension while acknowledging that clinicians regularly make interpretive errors. This expectation asymmetry reflects how people apply the machine heuristic, i.e., the rule of thumb that machines are more objective, perform tasks with greater precision, and handle information more securely than humans, thereby holding systems to higher reliability standards[41]. Furthermore, if a user observes system imprecision, then algorithm aversion may arise, i.e., "the tendency to prefer human judgments over algorithmic decisions even when it is suboptimal"[41]. In mental health contexts, this takes on particular significance and may be explained by the fact that genAI chatbots lack the therapeutic relationship, which has consistently been shown to drive significant change in clinical outcomes[42], allows human errors to be tolerated and repaired relationally[43]. Ultimately, young people still considered human-delivered care more desirable and authentic despite recognising instances of negative clinical experiences, reflecting relational cultural theory's insight that connection itself, not merely competent intervention, defines healing[44].

Relatedly, young people needed transparency of system functioning, knowledge base, reasoning, data practices, and accuracy. In mental healthcare, where errors can have serious consequences, such scrutiny is appropriate and aligns with established principles about transparency and interpretability for developing trustworthy and safe AI[45]. The connection between transparency and safety concerns in the workshop discussions, particularly regarding data breaches and unauthorised access, reflects research showing privacy is foundational to mental health help-seeking, given stigma[46].

While transparency was generally perceived as enabling empowerment and agency, young people recognised scenarios in which complete disclosure of AI-generated insights might be counterproductive, i.e., negative trajectories becoming self-fulfilling or triggering disengagement. Although this concern has been identified in the broader health literature relating to transparent information presentation practices and the risk of overwhelming clients[47], young people's perspective here suggests more salient risk to mental health outcomes and advocates a context-dependent rather than universal transparency approach. Additionally, young people's acknowledgement that decisions about risk detection implementation must be governed by clinical standards, not user preferences alone, reflects a nuanced understanding of when their agency should be steered by clinical responsibility[48] and user safety[33].

Through co-design, young people established Mia's role as a *guide*, joining the navigator, assessor, and educator roles across care journey touchpoints (see Table 2). Importantly, young people distinguished the guide role from companionship or friendship, emphasising that functioning as a friend would be inappropriate in mental health contexts where therapeutic boundaries and clinical relationships are paramount. This conception aligns with highly personalised and measurement-based care, positioning digital tools as support across care pathways[49]. Framed as a guide, Mia should draw out and retain mental health histories, reducing the burden and re-traumatisation from retelling experiences in fragmented care systems[50]; actively elicit information for developing personalised recommendations, consistent with evidence that assessing a client's strengths and resources is important for treatment planning[51]; and generate actionable self-management recommendations central to recovery[51]. Once a young person is linked to a service, Mia should gather updated information to revise recommendations, prepare users for sessions, conduct regular check-ins, and connect clients to immediate crisis support if needed. These functions address the persistent challenge of care continuity[52] and adhere to best practice, including between-session support that improves outcomes[53], immediate access to coping resources during distress that prevents escalation[54], and high quality, reliable information access which is crucial to support young people to self-advocate and make informed decisions[55].

**Table 2. Mia as a guide across care journey touchpoints.**

| Touchpoint | Navigator function | Assessor function | Educator function |
|---|---|---|---|
| **Pre-intake** | Explore service options without commitment | Complete self-screening and explore treatment options | Access psychoeducational content about mental health concerns |
| **Intake** | Orientation to service and book appointments | Complete assessments and forms, and create treatment and safety plans | Learn about intake process and treatment options |
| **Between sessions** | Reminders about appointments and homework | Access personalised self-care recommendations | Pre-session preparation materials |
| **Crisis** | Direct to appropriate care level and provide emergency contacts | Risk assessment and distress management recommendations | Evidence-based coping strategies |
| **Ongoing** | Query about service navigation | Track symptoms and mood, and receive updated recommendations | Query about treatments and build mental health literacy |

Young people's design requirements reflect broader youth digital mental health trends. Comprehensive user control mechanisms emphasise the importance of choice across modalities, consistent with evidence that personalisation improves engagement[56]. The distinction drawn between raw conversational data (private and user-controlled) and AI-generated insights (shareable for coordination) evidences an alternative data sharing approach for alleviating privacy-oversight tensions documented in youth mental health services, which typically follows a user-controlled, opt-in procedure[57]. Moreover, this approach appropriately balances young people's confidentiality with the clinical duty of care expected in a client-centred and empowering approach to youth mental health[58]. The two-part safety framing, i.e., distinguishing data safety from interaction safety, offers a useful structure for protective design going forward. The interaction safety expectations identified align with crisis intervention best practices[54] and user needs regarding genAI chatbots in mental health reported elsewhere[33], though current solutions often lack sophistication in handling mental health crises appropriately[59].

Successfully integrating genAI chatbots in youth mental health requires careful consideration and balancing of stakeholder needs and requirements, existing governance structures, and sociotechnical systems. Crucially, humanising genAI chatbots requires more than technical sophistication. Following responsible innovation principles[60], services must clearly communicate the role, capabilities, limitations, and appropriate expectations of these tools. Without such framing, gaps between expected and actual performance may undermine trust and risk disengagement. Integration challenges around data access expectations from multiple stakeholders with legitimate interests in young clients' wellbeing (e.g., guardians, educational institutions, and allied health services) that may conflict with youth privacy preferences reflect broader tensions in services[61]. Young people's nuanced perspective on privacy and safety challenges narratives positioning them as either privacy-unconcerned or categorically opposed to monitoring[62], underscoring the importance of considering these views to inform integration. Per sociotechnical systems theory, technology must fit operating contexts or risk disruption[63], and thus, without clarity about usage parameters, workflow integration, and governance frameworks, poorly integrated genAI chatbots risk disrupting care[64].

This work co-designed several concrete implications for the design, development, implementation, and governance of genAI chatbots in youth mental health. Design priorities include adaptive communication to manage expectations about system limitations, transparency and explainability in the presentation of potentially distressing information, granular privacy controls differentiating raw data from AI-generated insights, and potential risk response protocols governed by clinical standards. Development needs include diverse training data to reduce bias in AI-supported decision-making, robust user testing with young people to challenge systems, infrastructure supporting differential data access, and service system interoperability. Implementation requires transparent onboarding processes and positioning chatbots as complementary to human-delivered care. And governance requires clinician intervention pathways during crises, transparency about data practices, multi-stakeholder governance frameworks, and mechanisms for sustained lived experience input.

While the findings above have been discussed primarily in terms of young people's expectations and design requirements, it is critical to recognise these as substantive ethical concerns that transcend individual preferences and usability. Questions persist about whether genAI should be deployed in youth mental health, given the complexity of mental health, vulnerability of young people, and the fundamental role of human connection[65]. Young people's conditional acceptance advances but does not settle fundamental questions about dehumanisation risks, algorithmic decision-making requiring empathy and relational understanding, or the moral status of human-AI relationships in care. Yet, these perceptions provide some insight into how these technologies can be designed to avoid dehumanising care and well as delineating circumstances where they should and should not be deployed. Equity considerations carry particular moral weight. That is, genAI risks creating stratified care systems in which resourced individuals receive human support, amid global limitations on specialised service availability, while marginalised populations receive technological substitutes[66,67]. This raises concerns about justice when institutional efficiency benefits accrue to service systems while vulnerable young people experience diminished relational care. Yet, among them, those that are technically competent to utilise DMHTs are situated to gain huge advantages over those who do not in any circumstance due to better quality care associated with use of advanced tools[33], at least benefiting those with sufficient technology skills and access but still leaving behind those without, thus calling for improved digital access and literacy[68]. The transparency, privacy, and data governance requirements reported here are not merely preferences but ethical imperatives grounded in respect for autonomy, informed consent, and protection from harm[48]. Automated risk detection exemplifies complex ethical tensions between beneficence and autonomy in bioethics[69]. While young people in this study demonstrated sophisticated reasoning about proportionality (privacy for raw data and oversight for risk), implementing such a balance requires navigating competing values (i.e., protection versus liberty and a clinician's duty of care versus a client's self-determination) where stakeholders may disagree. Beyond being technical or design problems and solutions, these are ethical considerations that require an ethical framework, stakeholder deliberation, and ongoing reflection to address going forward.

Several limitations of this work should be noted. First, the study participants consisted of young people aged 18-30 willing to participate in mental health services research in the Australian context. As a result, the generalisability of the findings and conclusions drawn is inevitably shaped by local context, participant ages, and the absence of certain voices and perspectives in this study, particularly those most marginalised due to associated stigma or aged 18 years or younger. Yet, drawing on the strengths of reflexive thematic analysis, the qualitative analysis engaged a diverse group of researchers (including lived experience researchers) at multiple points to contribute a range of perspectives and to develop a comprehensive examination of young people's perspectives and recommendations. Second, all workshops were conducted and recorded remotely, and thus, in some instances, the quality and accuracy of the transcripts were affected by internet connectivity issues. Third, despite efforts to enhance co-design inclusiveness (i.e., enabling remote participation, engaging lived experience researchers, and purpose sampling), broaden interpretive perspectives (i.e., engaging multiple, diverse researchers in data collection and analysis), and prevent a reductionist framing (i.e., featuring direct participant quotations in reporting), such limitations cannot be fully eliminated. Fourth, the primary focus on a single genAI chatbot (Mia) in a single context limits transferability, though the identified principles and implications are likely broader in relevance. Finally, participants engaged with functioning prototypes at one time point rather than final products over an extended period of user testing, and thus, user experiences may differ from anticipated ones.

Overall, young people reported the value of genAI chatbots for consumers in youth mental health, contingent on humanising AI without dehumanising care, transparency, strategic positioning and service integration, and balanced control with safety for sustained engagement. Their nuanced perspectives and recommendations reflect the complexity of introducing genAI into domains defined by vulnerability, trust, and relationships, as well as a careful balancing of competing priorities among stakeholders. The challenge ahead for integrating genAI chatbots into youth mental health services extends beyond technical capability to encompass managing expectations, framing appropriate roles, and ensuring technological pursuit does not devalue human relationships defining meaningful care. Realising potential requires centring youth voices as ongoing partners in design, development, implementation and governance, honestly grappling with inherent tensions, and prioritising wellbeing over technological enthusiasm or cost savings. Co-designed insights reported here provide a foundation going forward, but translating principles into ethical, effective, equitable systems that advance highly personalised and measurement based youth mental healthcare requires ongoing, meaningful work to appreciate sociotechnical systems and embed lived experience engagement.

## 4. Methods

### 4.1. Conversational generative artificial intelligence agent development

We developed a prototype genAI chatbot and agentic system for youth mental health, the Mental health Intelligence Agent (Mia)[35], and made it available to young people who participated in this study. Mia was prototyped with insights from internal stakeholders representing expertise in psychiatry, general and clinical psychology, youth mental health service management, technology development, and human-computer interaction[35]. Primarily, Mia utilises LLMs, a curated database, and a reasoning engine to scale expertise with specific knowledge about youth mental health related to the Brain and Mind Centre Youth Model of Care[51,70]. The database consists of evidence-based literature, expert knowledge, and best practice clinical guidelines. In use, Mia interprets user input concerning a young person's mental health information (e.g., previous and current symptoms, case formulation, support, and service engagement) and other associated, contextual factors; highlights missing critical information and produces structured summaries and timelines; elicits information relating to critical signs to assess for in young people, principled multidimensional assessment, and confidence threshold criteria for developing a treatment plan; generates prioritisation guidance; and presents highly personalised and context-driven guideline-linked suggestions for advising treatment options, future assessment, and service navigation. Outputs are advisory and explainable, i.e., for each output, the inputs considered, logic/criteria, and citations are shown. Following an iterative and responsive co-design approach, regular design sprints were conducted throughout the study to incorporate evolving insights from young people into subsequent Mia prototype versions, which were then shown to new participants. More details on specific features and functionality of Mia, as well as the user interface, are provided throughout the presentation of findings.

## 4.2. Study design

This study engaged young people in online workshops to inform the development of Mia following a co-design methodology, centring on transforming it from a health professional tool into a solution for consumers. We expanded the focus of the workshops to investigate participants' perceptions and recommendations about genAI chatbots in youth mental health generally and Mia specifically. Co-design is frequently employed to collect qualitative insights and stakeholder design considerations regarding the introduction of various digital health technologies in different socio-technical contexts (e.g., health and wellbeing apps[71,72]).

## 4.3. Participants, recruitment, and sampling

Participants were recruited via snowball and purposive sampling, as part of a series of research projects addressing youth mental health. Participants comprised young people aged 18-30. Existing research and community-based networks were engaged to support the distribution of recruitment materials among their networks. Additionally, individuals who participated in prior or parallel research projects conducted by the wider research team and gave their informed consent to be contacted directly by the research team for future research conducted by the research team were sent recruitment materials for this study. Individuals responded directly to the research team if they wished to take part, and were thereafter sent a participant information sheet and consent form via an online survey (via Qualtrics) to record their informed consent.

## 4.4. Data collection

Data were collected online via Zoom and Qualtrics between March and August 2025 using participant workshops and surveys. Two types of workshops were held. First, a series of five conceptual workshops, focusing on exploring young people's perceptions and developing recommendations regarding genAI chatbots in youth mental health generally and Mia specifically, with a live, interactive demonstration of Mia. Second, a set of four user testing workshops that involved a period of free-form user testing and a think-aloud component, followed by a semi-structured discussion, which included brainstorming and voting activities. All workshops were audio-video recorded and transcribed verbatim with the assistance of Otter.ai, an automated transcription platform. Surveys were conducted prior to workshop participation and collected participants' demographic data. All data were anonymised prior to analysis. Data was collected by AP, HML, and FI.

## 4.5. Data analysis

Data were analysed following inductive reflexive thematic analysis to develop and report themes based on patterns of shared meaning in the workshop data, without a pre-determined set of themes[73,74]. Following reflexive thematic analysis, six iterative analytic phases were completed: data familiarisation, generating initial codes, generating candidate themes, reviewing candidate themes and establishing final themes, defining and naming final themes, and producing the report[36].

To start, AP and HML independently reviewed one workshop transcript and generated and shared initial codes. Thereafter, AP coded the remaining data. At the midway point, HML independently coded a second workshop transcript to contribute new, evolving insights and enrich the ongoing data analysis. To generate, review, and finalise the themes, the wider research team was engaged in discussion. This added key insights from expertise in lived experience, psychiatry, general and clinical psychology, and human-computer interaction. NVivo 15 was utilised to organise and code the workshop data. Descriptive statistics were employed to develop frequency analysis of the survey data.

### 4.6. Ethics approval

This research was approved by the University of Sydney Human Research Ethics Committee (Protocol no: 2021/680). This study was performed in line with the principles of the Declaration of Helsinki and according to the guidelines of the National Statement on Ethical Conduct in Human Research and the Australian Code for the Responsible Conduct of Research.

## Declarations

### Data availability

Raw data are not available to the public due to data protection and privacy reasons. Anonymised metadata (e.g., code frequency) that support the findings are available from the corresponding author upon reasonable request.

### Funding

This work was supported by a philanthropic funding donor affected by mental health who wishes to remain anonymous. IBH was supported by an NHMRC L3 Investigator Grant (GNT2016346). WC was supported by the Australian Government Research Training Program (RTP) Scholarship. FI was supported by an NHMRC EL1 Investigator Grant (GNT2018157).

### Author contributions

**A.P.:** Conceptualization, Methodology, Formal analysis, Investigation, Data Curation, Writing–Original Draft, Writing–Review & Editing, Visualization. **I.B.H.:** Resources, Writing–Review & Editing, Supervision, Project administration, Funding acquisition. **C.G.:** Formal analysis, Writing–Review & Editing. **Z.H.:** Writing–Review & Editing. **W.C.:** Writing–Review & Editing. **E.E-A.:** Software, Writing–Review & Editing. **J.R.:** Software, Writing–Review & Editing. **E.M.S.:** Writing–Review & Editing. **F.I.:** Conceptualization, Methodology, Formal analysis, Investigation, Resources, Writing–Original Draft, Writing–Review & Editing, Supervision, Project administration. **H.M.L.:** Conceptualization, Methodology, Formal analysis, Investigation, Resources, Writing–Original Draft, Writing–Review & Editing, Supervision, Project administration.

**Competing interests**

I.B.H. is a Professor of Psychiatry and the Co-Director of Health and Policy, Brain and Mind Centre, University of Sydney. He has led major public health and health service development in Australia, particularly focusing on early intervention for young people with depression, suicidal thoughts and behaviours and complex mood disorders. He is active in the development through codesign, implementation and continuous evaluation of new health information and personal monitoring technologies to drive highly-personalised and measurement-based care. He holds a 3.2% equity share in Innowell Pty Ltd that is focused on digital transformation of mental health services. E.M.S. is a Principal Research Fellow at the Brain and Mind Centre, University of Sydney, a Consultant Psychiatrist and Adjunct Clinical Professor at the School of Medicine, University of Notre Dame. She previously served as the Discipline Leader for Adult Mental Health at Notre Dame until January 2025. In addition, she is a member of Medibank's Medical and Mental Health Reference Groups. A/Prof Scott has also delivered educational seminars on the clinical management of depressive disorders, receiving honoraria from pharmaceutical companies including Servier, Janssen, and Eli Lilly. Moreover, she has contributed to a national advisory board for Pfizer's antidepressant Pristiq and served as the National Coordinator for an antidepressant trial sponsored by Servier. A.P., C.G., Z.H., W.C., E.E-A., J.R., E.M.S., F.I., and H.M.L. have nothing to disclose.

**Figure 3. Original Mental health Intelligence Agent (Mia) user interface intended for health professionals, utilised for early stages of user testing.**

**a. Home screen.**

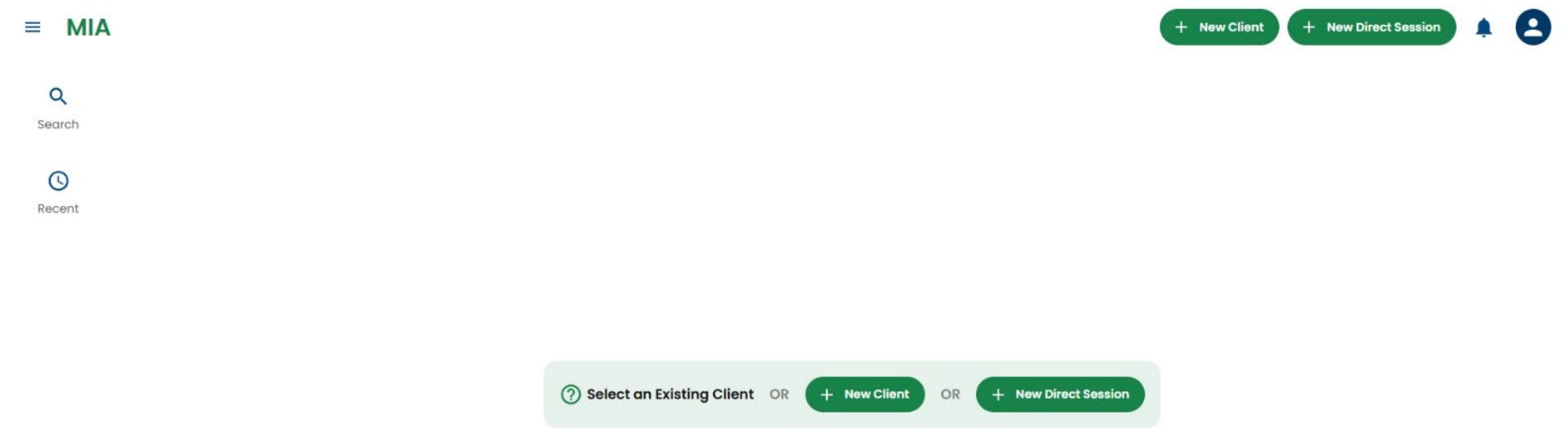

**b. First user input screen.**

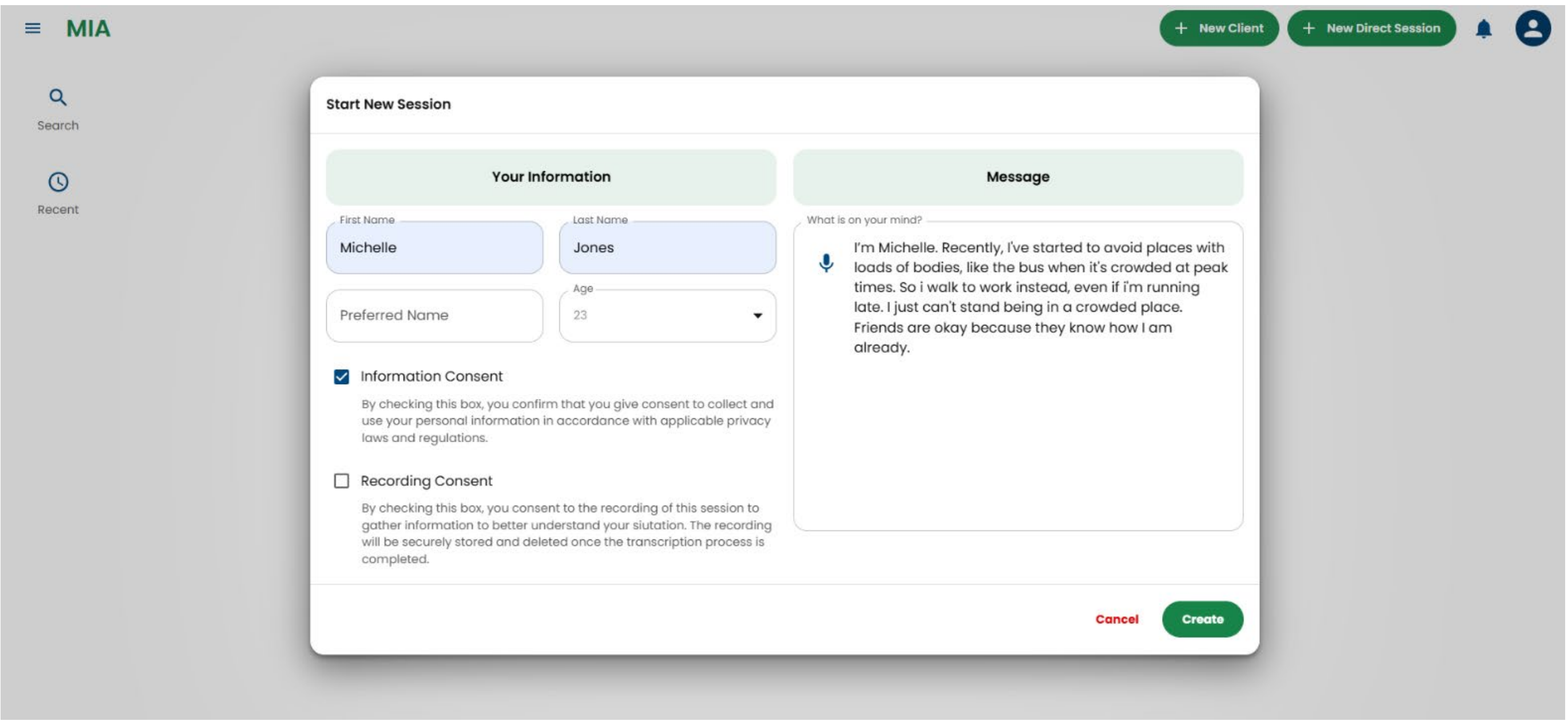

**c. Main screen with persistent chat panel (left) showing ongoing messaging between the user and system, and tabbed information panel (right) showing the structure of information to be collected and insights to be generated.**

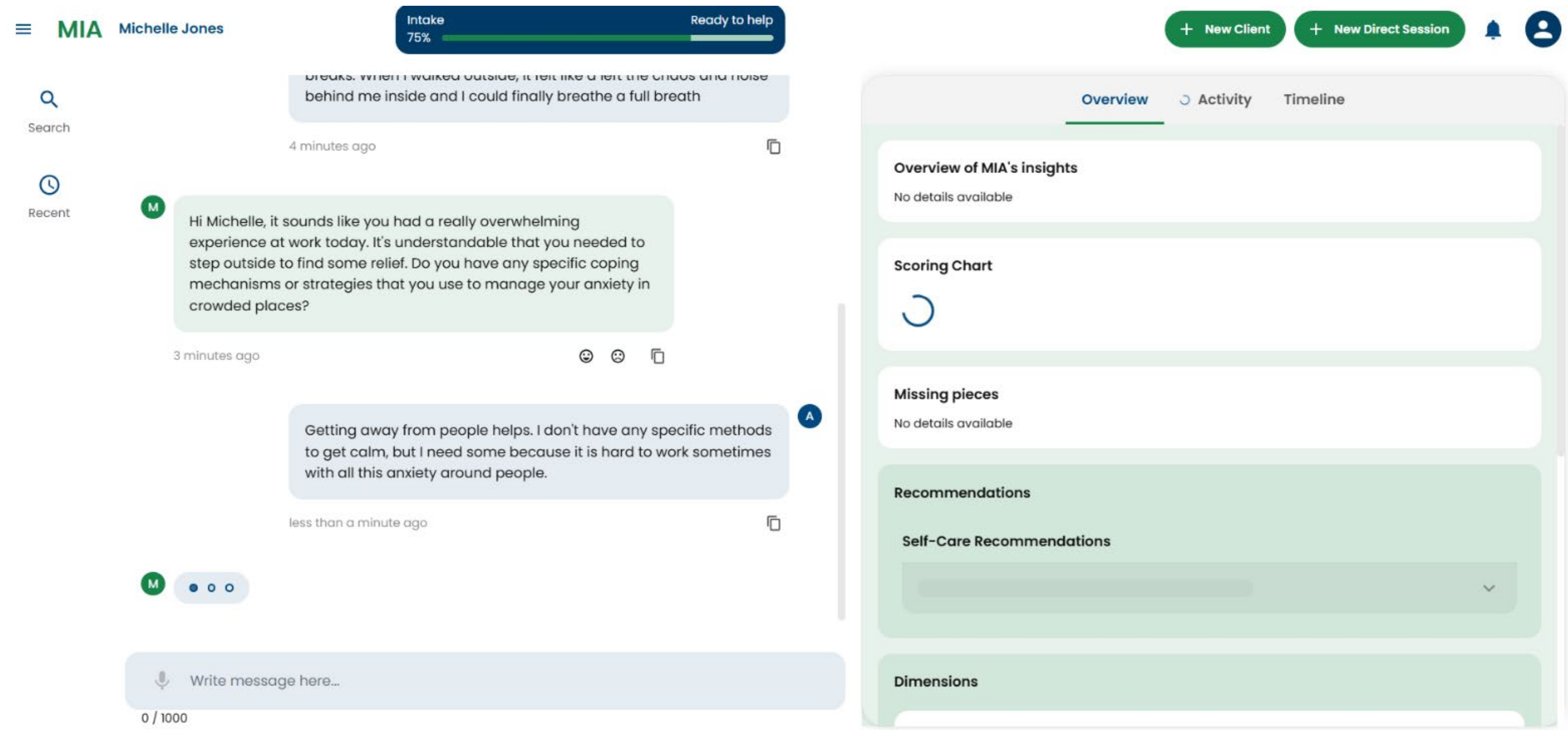

**d. Information threshold reached and thus insights generated, showing recommendations and information organised by mental health dimensions (composite image due to the information panel being scrollable).**

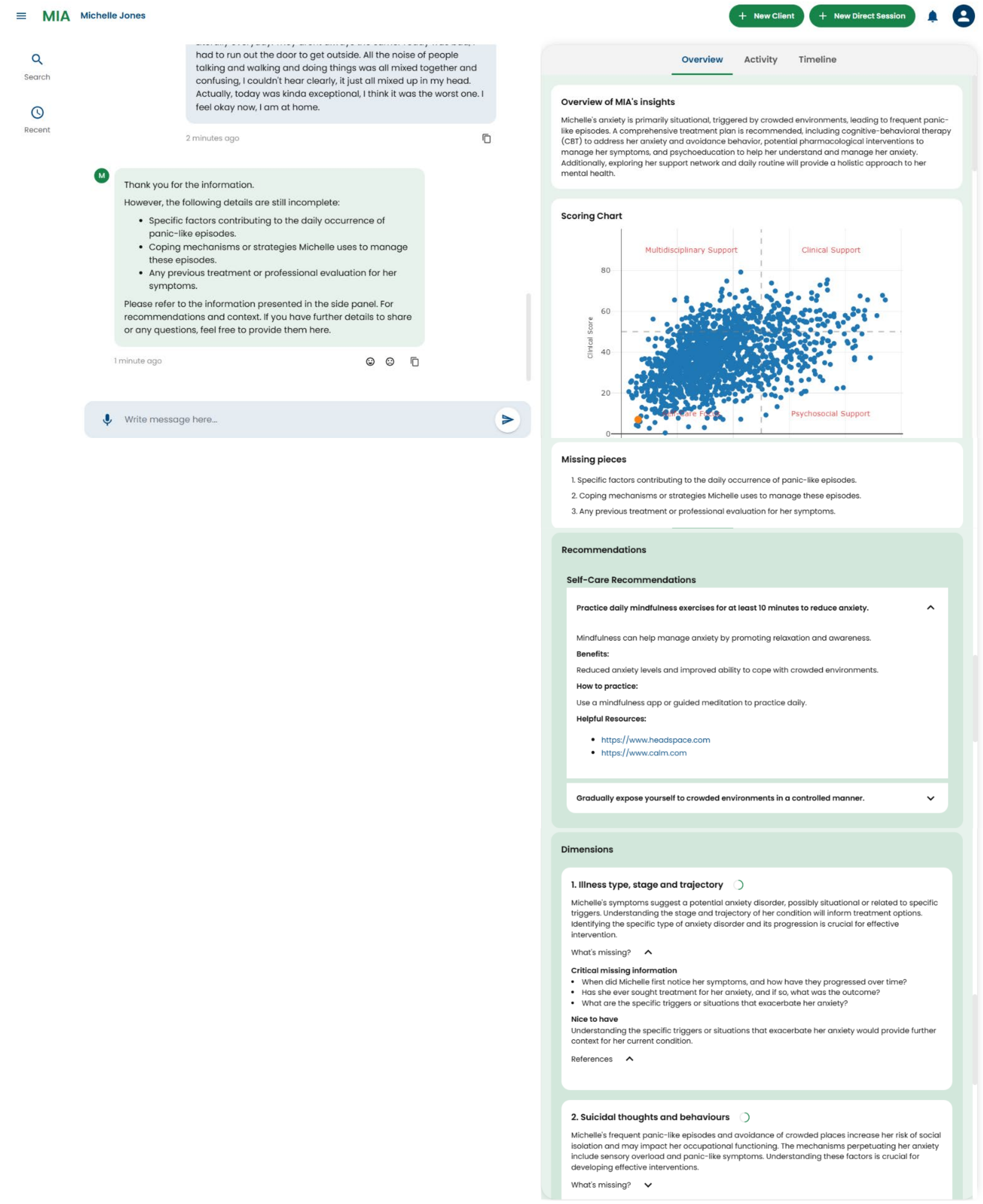

**e. Timeline tab, showing a suggested event to add to the user's information automatically derived from user input.**

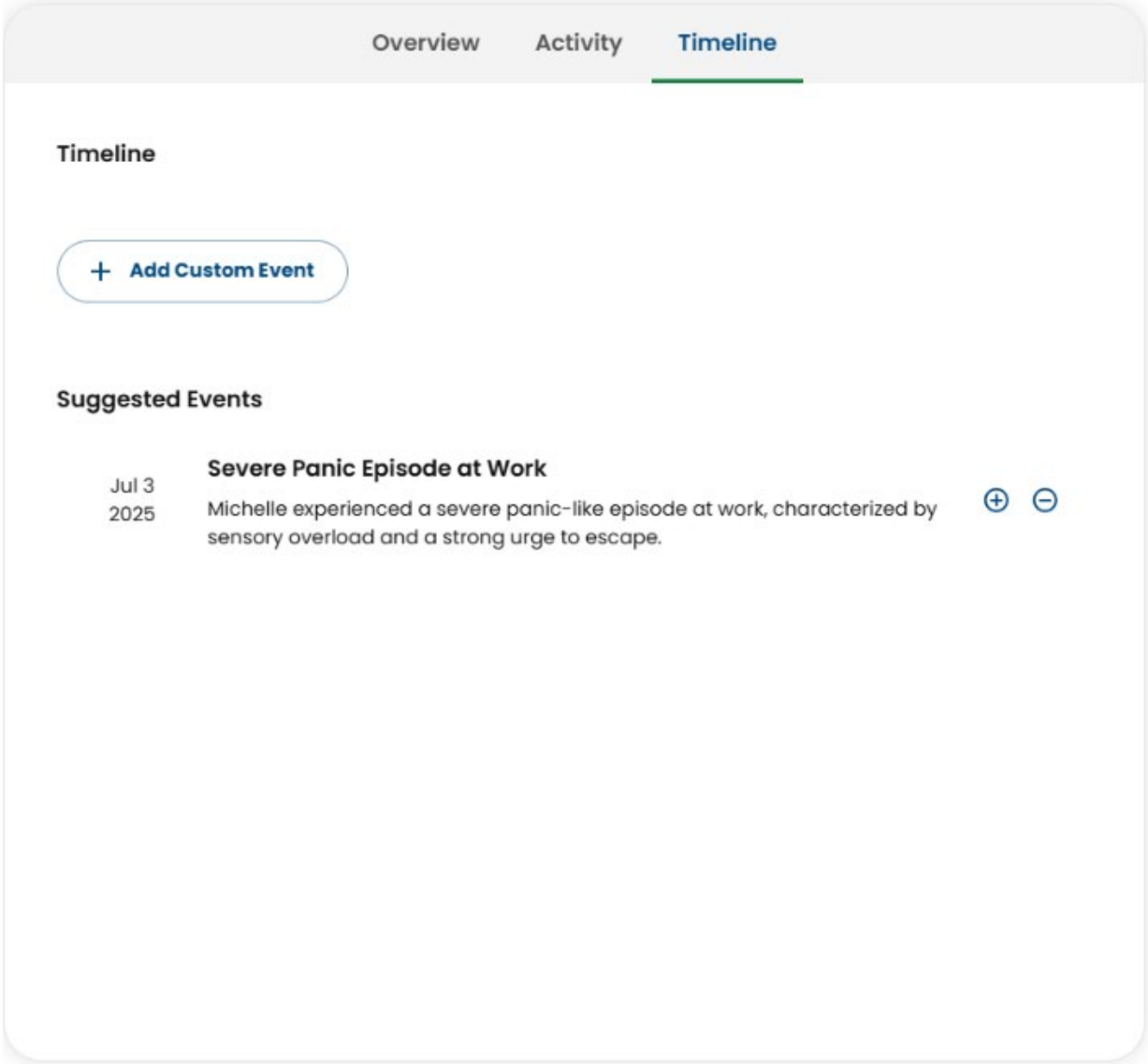

**Figure 4. Revised Mental health Intelligence Agent (Mia) user interface updated for young people based on feedback, used for later stages of user testing.**

**a. Home screen, with added introductory panels featuring user disclaimers and system operations information.**

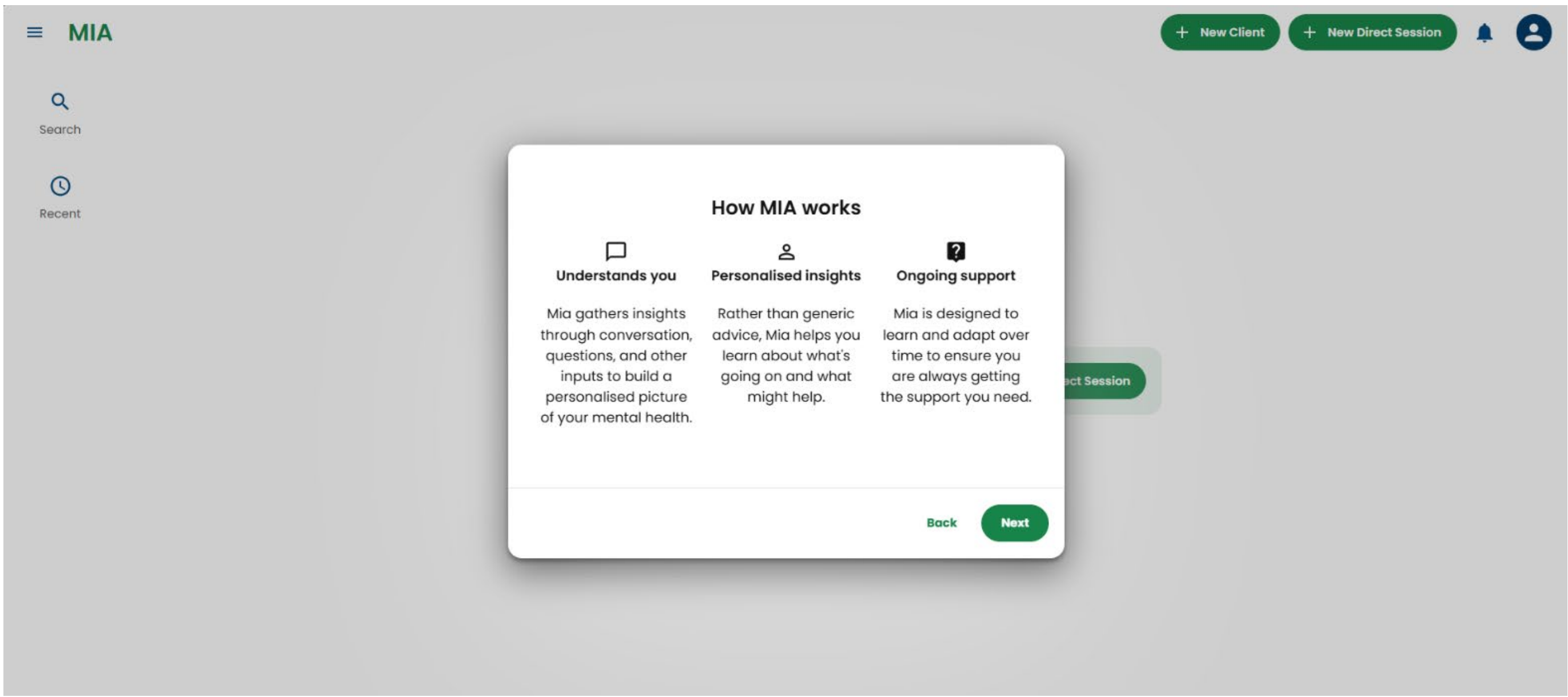

**b. First user input screen.**

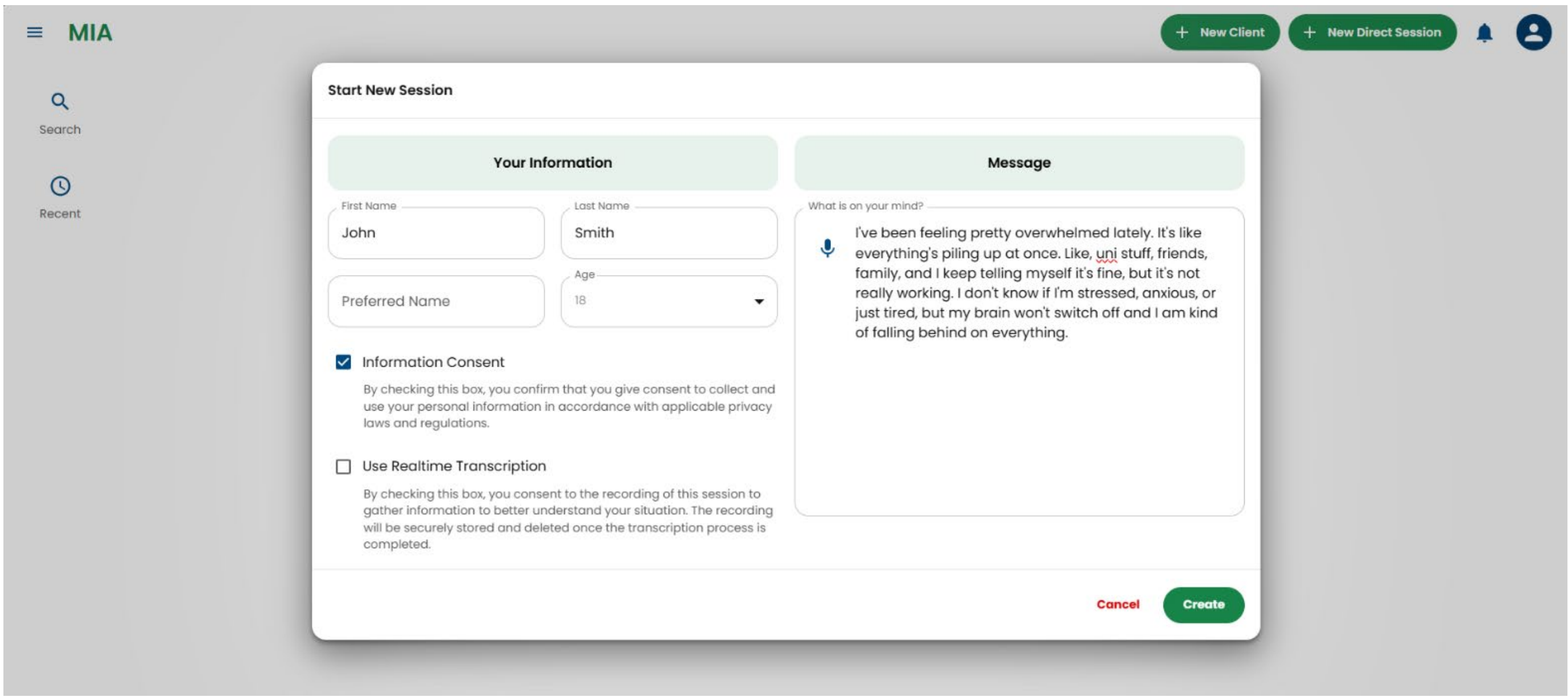

**c. Main screen, with revised information panel (right) featuring an information completeness visualisation to communicate how much information the system has about the user and in-progress assessment insights listing what the system knows about the user, what topics it is currently exploring with the user, and the system's working assumptions about the user (composite image).**

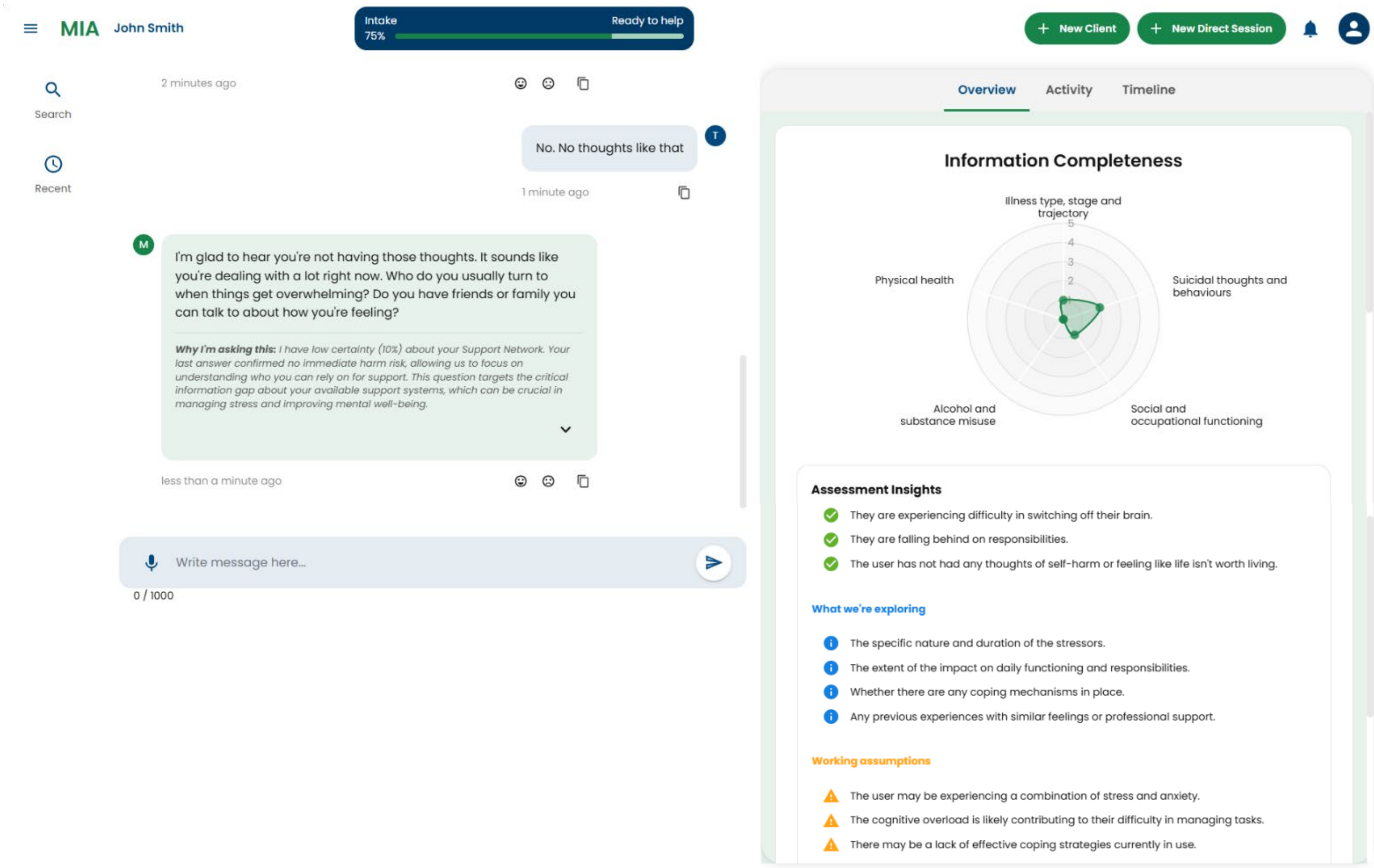

**d. Information threshold reached and thus insights generated, showing revised information panel featuring personal experiments (previously, recommendations), missing pieces and service needs index visualisation (shown as a popout here due to space limitations). In-progress assessment insights are reframed as missing pieces. Information completeness hidden from user (composite image).**

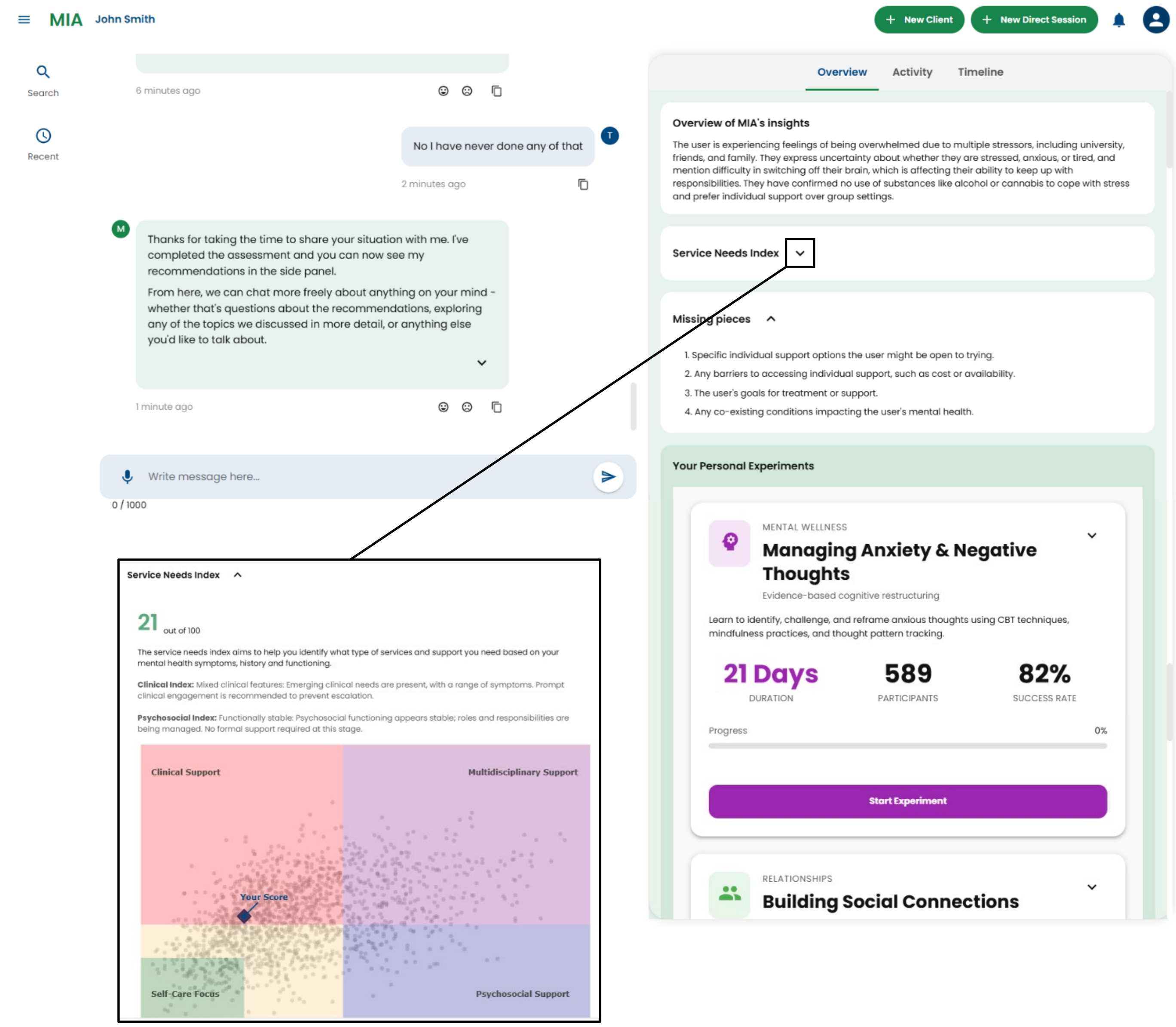